# Delayed Feedback and Chaos on the Driven Diode-Terminated Transmission Line

Vassili Demergis[1], Alexander Glasser[2], Marshal Miller[2],
Thomas M. Antonsen Jr.[2,3], Edward Ott[2,3], Steven M. Anlage[1]

1. Center for Superconductivity Research, Department of Physics, University of Maryland, College Park, Maryland 20742-4111, USA
2. Institute for Research in Electronics and Applied Physics, University of Maryland, College Park, Maryland 20742-3511, USA
3. Department of Electrical and Computer Engineering, University of Maryland, College Park, Maryland 20742-3285, USA

## Abstract

A simple model of a distributed, non-linear circuit that produces chaos at GHz frequencies is introduced and tested experimentally. The model circuit is a driven diode-terminated transmission line with the transmission line impedance mismatched to that of the source. Experiments were performed with sinusoidal driving frequencies of 10 MHz to 1.2 GHz, driving powers of -30 to +50 dBm and transmission line delay times from 3 to 20 ns. Diode reverse recovery times ranged from 4 to 100 ns. As a result of many experiments, it was found that chaotic behavior was strongly dependent on the unbiased, small signal reactance of the system as seen by the driving source, and influenced by an applied DC voltage-bias across the diode. In the experiments that showed period-doubling and / or chaos, the reverse recovery times of the diodes were on the order of both the driving period and the delay time of the circuits. Comparisons between theory and experiment are in general agreement. Chaos produced with a driving frequency of 1.105 GHz has been observed experimentally.

**Controlled GHz-scale chaos is a new frontier for both science and technology. Methods of generating and understanding such high frequency chaos are just beginning to be studied. The approach taken here is an extension of the driven lumped element resistor-inductor-diode circuit chaos and focuses on the importance of time scales and the nonlinear capacitance of the diode. This is in contrast with studies which focus on the nonlinear current-voltage curve of the diode. With the increasing operating speeds of computers, there is an increasing demand to understand the dynamics of high frequency signals on computer bus lines. It has been shown that chaotic instabilities could arise at GHz frequencies from the electrostatic discharge protection circuits commonly found surrounding most computer components. Through an understanding of the general system, it may be possible to avoid such instabilities. In addition, systems that can create chaos at GHz frequencies may enable high bandwidth communication technologies using synchronized chaotic oscillations.**

## Introduction

The p/n junction is well-known for its nonlinear properties – from voltage dependent capacitance and nonlinear current-voltage characteristics, to the memory and reverse recovery effects of the minority carriers. These nonlinearities make p/n junction diodes ideal for creating circuits which demonstrate chaotic behavior and yield chaotic waveforms.

The nonlinear dynamics of the driven, lumped-element series resistor-inductor-diode (RLD) circuit have been thoroughly studied, and the key physical phenomena giving rise to chaos in this classic p-n junction system are essentially understood.[1-11] It has been established that the nonlinear capacitance, as well as the reverse recovery time of the diode, both play important roles for the development of chaos.[3,8-11]

It has been found that chaos and period doubling in the driven, lumped-element, series RLD circuit is best understood by examining the time scales in the problem. First, period doubling and chaos occur most readily for driving frequencies $f$ near the resonant frequency of the system: $f_0 \approx 1/\left(2\pi\sqrt{LC_j(0)}\right)$, where $L$ is the value of the inductor element and $C_j(0)$ is the junction capacitance of the diode measured at 0 volts. It was also found that the reverse recovery time of the diode ($\tau_{RR}$) should be on the order of the driving period:[10] $\tau_{RR} \approx 1/f$. There are many non-linearities of the reverse recovery time itself. For example, the value of $R$ in the circuit can play a role in determining the effective $\tau_{RR}$ because the reverse recovery process involves the discharge of the diode back through the rest of the circuit.[10] Reverse recovery effects are not explicitly considered in most numerical simulations of the driven RLD circuit. Instead, most authors consider a voltage-dependent capacitance, sometimes as simple as a two-state capacitor.[6,12] However, these models retain a remnant of reverse recovery effects through the RC time constants for the discharge of the capacitor which mimic the charge-storage reverse recovery dynamics of the diode.[10]

Generally, the reverse recovery time of packaged diodes is in the range of about $10^{-8}$ s to $10^{-5}$ s, and consequently the frequencies associated with generating chaos are low (kHz to MHz range). There has been a growing need to understand the origins of chaos in electrical circuits at higher frequencies. In modern computer electronics, operational frequencies of devices are in the 0.1 to 10 GHz range. It is in this range that we intend to generate and understand chaos.

The potential for nonlinearity and chaos in modern computer circuits has already been demonstrated. Electrostatic discharge (ESD) protection diodes are integrated into almost all electronic devices as a means of protecting the logic circuits from potentially destructive outside signals and high-voltage discharges. Parasitic inductance and capacitance associated with the diodes and their packaging can create nonlinear RLD circuits that resonate at GHz frequencies. It has been shown through out-of-band excitation that these systems can be driven into nonlinear behavior and possibly chaos at GHz frequencies.[13-15] These chaotic signals could then be transmitted into the primary circuitry and disrupt, or possibly damage, the device.

Our model (discussed below) may simulate computer interconnect transmission lines used to drive logic circuits. The ESD protection diodes effectively terminate the transmission line, and reflected and re-reflected waves will provide a complicated delayed feedback to the p/n junction. In addition, RF signals will be rectified by the p/n junction, giving rise to a DC voltage bias and to even more complicated nonlinear behavior.[11]

There is also a practical need to generate chaotic microwave signals. With the discovery of chaotic control[14] and synchronization,[17,18] uses of chaotic signals for communication,[19] including low probability of detection and low probability of intercept data transfer,[18,20] have become more feasible. Since chaos is seemingly noise-like but deterministic, it is possible to hide data streams within this chaotic "noise" and decipher the signals on the receiving end using chaotic synchronization techniques. In addition, even when low probability of interception is not an issue, communicating with chaotically generated signals can have other advantages, such as transmitter design simplicity[19] and relay power efficiency.[21] It is the purpose of the present paper to present results on chaos generation in a transmission line diode circuit. Because a transmission line, rather than lumped circuit elements is used, the scheme we investigate is capable of producing chaos with relatively short time scales, thus possibly accommodating larger information transfer rates.[22]

Our approach is to develop a generalized version of the driven RLD circuit to generate chaos at RF and microwave frequencies. Essentially, the inductor is replaced with a distributed transmission line, thus introducing a new memory/feedback mechanism on the diode as well as a new time scale.

There has been some prior work related to that discussed here. A version of Chua's circuit has been analyzed where the passive, linear, inductor / capacitor portion has been removed and replaced by a shorted, lossless transmission line to provide a time delayed feedback.[23] To simplify the problem, the capacitor in parallel with Chua's diode was removed. Essentially, the circuit described becomes a lossless transmission line terminated on one end by a short and on the other end by a resistor and Chua's diode in series.[23]

There has also been work on a self-oscillating transmission line system terminated on one end by a diode and on the other by a negative resistor. Numerical results show period doubling and chaos when the magnitude of the reflection coefficient, ρ, at the negative resistor boundary is greater than 1.[24] In other words, this system uses internal amplification of the signal to produce period doubling and chaos. Through numerical analysis, period doubling can be seen through varying values of ρ.[24] An experimental study of this system has been conducted, and direct comparisons have been made between the theoretical and experimental results.[25] Further study has been conducted on delayed feedback systems[26] and the transmission line limit of the Rössler system.[27,28]

Other work has been done to experimentally study the spatial-temporal properties of a passive nonlinear resonator in the case of high external driving power.[29] In this case, a resonator cavity was constructed which was terminated by a short on one end and by a varactor diode on the other end. A signal generator connected to a loop antenna drives the resonator near the shorted end. Measurements of the mean-square diode voltage as a function of frequency and power were made, and it was found that, as driving power increased, the resonances became distorted and hysteresis was observed.[29] Time domain waveforms of the diode voltage and electric field measurements along the length of the resonator were also measured, and a phase diagram of "stochastic oscillations" as a function of power and frequency was made.

The model of the time-delayed, distributed diode problem that will be discussed in this paper is similar to, yet different from, the above approaches. Here, the transmission-line driven diode will be treated as a generalization of the classic, lumped RLD circuit.[1-11] The addition of a transmission line introduces a new time scale associated with the delayed feedback of the reflected and re-reflected waves. Another unique feature is the inclusion of a variable impedance mismatch between the generator and the transmission line. Thus the value of the reflection coefficient ρ can be changed in a controlled manner. This system also serves as a realistic model of the ESD diode problem in modern computer circuits and opens up new opportunities to generate chaos at GHz frequencies with relatively simple and ubiquitous systems.

## Model

The model in Fig. 1 shows a harmonically driven, distributed, lossless transmission line terminated with a diode. It is important to note that the impedances of the source and transmission line are mismatched so that the wave reflected off the diode can be linearly re-reflected at the source boundary. The three governing equations involving the incident ($V_{inc}$), reflected ($V_{ref}$) and total ($V$) voltage on the diode can be written down:

$$2V_{inc}(t) = V(t) + Z_0\left[gV(t) + \frac{d}{dt}Q\right] \quad \textbf{(1a)}$$

$$V_{ref}(t) = V(t) - V_{inc}(t) \quad \textbf{(1b)}$$

$$V_{inc}(t) = \rho V_{ref}(t-2T) + \tau V_g(t-T) \quad \textbf{(1c)}$$

Here, $Z_0$ is the characteristic impedance of the transmission line, $V$ is the voltage across the diode, $V_g$ is the voltage generated by the source, $V_{inc}$ ($V_{ref}$) is the voltage of the incident (reflected) wave at the diode, $\tau$ is the voltage transmission coefficient from the source into the transmission line, $\rho$ is the voltage reflection coefficient from the transmission line off the source boundary, $Q$ is the charge on the diode p-n junction, and $g$ is the conductance of the diode. In what follows, we assume that $Q$ at time $t$ is a nonlinear function of the voltage at time $t$, $Q = Q(V(t))$, i.e., there is no past history dependence of $Q$. The diode is modeled as a nonlinear capacitor and linear resistor in parallel. It is known that the nonlinearity of the shunt resistor is not needed for generating chaos,[6] and the resistor will, in our model, be treated as linear with constant conductance value $g$ in Eqn. 1 ($g = 1/R$).

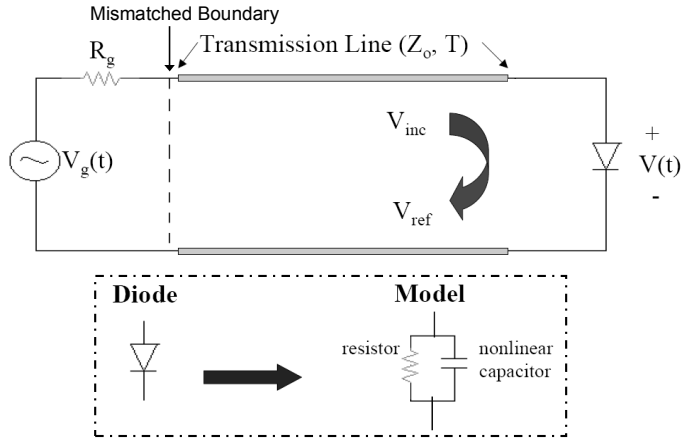

Fig. 1: Schematic diagram of the model of the driven, lossless, diode-terminated transmission line. Shown underneath is the equivalent electrical schematic model of the diode.

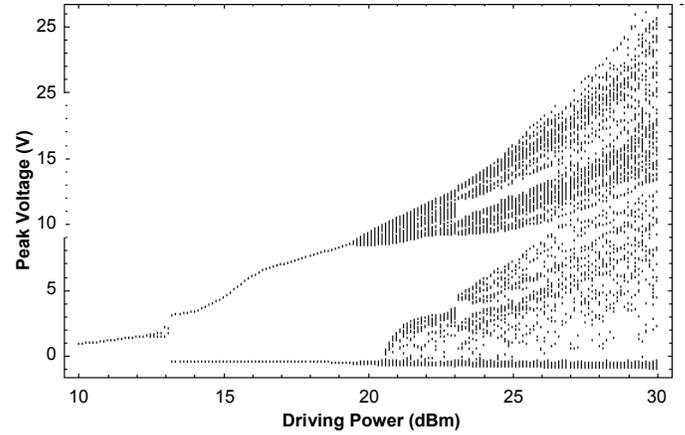

Fig. 2: Numerical bifurcation diagram of the model shown in Fig. 1 and Eq. (2) with parameters: $f$=85MHz, $Z_0$=50Ω, $\rho$=$\tau$=0.85, T=8.45ns, R=200Ω, $C_f$=1.5nF, $C_r$=1.0pF, $V_f$=0.6V, $V_{gap}$=0.1.

Eqn. 1a describes the voltage across the diode based on the incident voltage wave and the diode model. Eqn. 1b states that the diode voltage is the sum of incident and reflected waves. Eqn. 1c describes the new incident wave as the sum of the reflected wave and the source-generated wave multiplied by reflection and transmission coefficients respectively. Note that a time scale has been introduced to the system – the one-way delay time, $T$, of the transmission line.

The three equations can be used to solve for the voltage drop across the diode. The source is modeled as a sinusoidal driving signal, $V_g(t) = V_g Cos[\omega t]$ where $\omega$ is the driving frequency. We also rewrite $Q(V(t))$ as $V(t)C(V(t))$, representing the $Q(V)$ relation through a nonlinear capacitance. The equations can be combined into a single, first-order, delay differential equation:

$$\frac{d}{dt}V(t) = -\frac{(1+Z_0 g)}{Z_0 C(V(t))}V(t) + \frac{\rho(1-Z_0 g)}{Z_0 C(V(t))}V(t-2T) - \frac{\rho C(V(t-2T))}{C(V(t))}\frac{d}{dt}V(t-2T) + \frac{2\tau V_g}{Z_0 C(V(t))}Cos[\omega(t-T)]$$
(2)

We solve this equation using the fourth order Runge-Kutta method and analyze the resulting time domain waveforms through bifurcation diagrams, frequency-power spectra, and phase diagrams.

There are several key parameters in this model: The frequency of the driving signal $\omega$, the one-way delay time $T$, and the transmission and reflection coefficients ($\tau$ and $\rho$ respectively) at the boundary between the source and the transmission line. The diode capacitance is modeled by a hyperbolic tangent function (to mimic a step function between two values):

$$C(V) = \frac{C_f + C_r}{2} + \frac{C_f - C_r}{2}\tanh\left(\frac{V(t)-V_f}{V_{gap}}\right) \quad (3)$$

Here, $V_f$ is the voltage at which the capacitance switches from its low value ($C_r$) to its high value ($C_f$), and $V_{gap}$ controls how fast this switching occurs.

The model shows period doubling and chaos at high drive amplitude levels for realistic parameter values. Fig. 2 shows a bifurcation diagram constructed through numerical calculation for the case of an 8.45ns transmission line delay time with a mismatch giving $\rho = \tau = 0.85$ with the source. A few general observations can be made from the numerical simulations. First, chaos and period doubling occur more readily when the magnitude of $\rho$ is close to 1. This illustrates the influence of the reflected wave off the source boundary. Also, the larger the value of $\tau$, the lower the driving voltage required for the first period doubling transition. It was also seen that period doubling and chaotic behavior were observed in regular intervals as a function of driving frequency with a fixed transmission line delay and all other system parameters fixed. The frequencies at which chaos was readily observed correspond with frequency ranges that induce a positive reactance in the system as seen by the source. This will be discussed in further detail below.

## Experiment

Two variations of this model were constructed for the experimental tests. Fig. 3(a) shows the most basic design of a source driving an impedance-mismatched transmission line terminated by a diode. This is known as the "Partially Reflecting" experiment. Note that compared to Fig. 1, additional microwave components were added for isolation and measurement purposes.

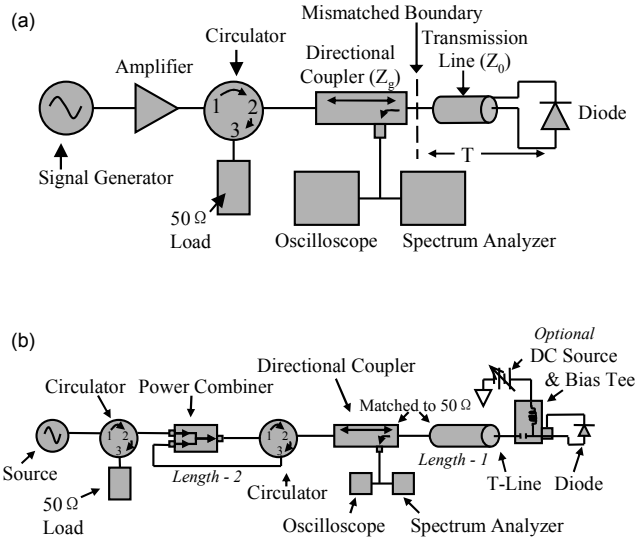

**Fig. 3: (a)** Schematic diagram of the Partially-Reflecting experimental setup. **(b)** Schematic diagram of the Bent-Pipe experimental design with optional DC biasing components.

Traditional probe measurements of diode voltage cannot be made at the frequencies of interest without corrupting the circuit properties. Instead, the reflected wave from the diode/transmission line is measured using the directional coupler so that we do not directly interfere with the system.

Fig. 3(b) shows a modified version of the basic experiment. It was found from numerical and PSPICE simulations of the circuit that period doubling and chaos occur more readily when both the magnitude of the transmission and reflection coefficients became close to 1. In Fig. 3(a), the driving signal is inserted using 50Ω coaxial cable into a mismatched transmission line coaxial cable. Simple wave physics considerations show that the sum of the absolute magnitudes of the reflection and transmission coefficients must be 1, and so they cannot both be close to 1 simultaneously. The design in Fig. 3(b) allows the reflection and transmission coefficients of the model equations to both be close to 1 first by matching the impedances of all the transmission lines to 50Ω. Then, rather than reflect a signal off a mismatched boundary, a circulator and power combiner work in tandem to turn the reflected signal and send it back to the diode. Now, the effective values of $\rho$ and $\tau$ can be made close to 1 in magnitude. This is known as the "Bent-Pipe" experiment. Table I summarizes the parameter values used in the experiments.

**Table I: Parameter values and components used in Partially Reflecting and Bent-Pipe Experiments**

| Parameters | Partially-Reflecting Experiment | Bent-Pipe Experiment |
|---|---|---|
| Drive Frequency | 50 MHz → 1.2 GHz | 50 MHz → 1.2 GHz |
| Driving Powers | -30 dBm → +50 dBm | -30 dBm → +50 dBm |
| Delay Times (ns) | 8.6, 17.3 | 3.0, 3.5, 3.9, 4.1, 4.4, 5.5, 7 |
| Diode $\tau_{rr}$ (ns) | 4, 5, 30, 35, 100+ | 4, 5, 30, 35, 100+ |

In the Partially Reflecting experiment (Fig 3(a)), the signal is generated by the source and passes through port one of the circulator and out port two, then passes through the directional coupler. The wave goes through the mismatched boundary into the transmission line and is reflected off the diode. The wave travels back down the transmission line and is both reflected off the mismatched boundary and transmitted through it with coefficients $\rho$ and $\tau$ respectively. The transmitted part is sent through the directional coupler and a portion of that wave (-30dB or -10dB) is measured by both a spectrum analyzer (Agilent 4396B) and an oscilloscope (Agilent Infiniium DSO81204A or Tektronix TDS3052). It is important to note that for high frequencies (nearing 1 GHz) the oscilloscopes are being pushed to their limits and therefore the signal to noise ratio may be poor. Therefore, frequency-power spectra given by the spectrum analyzer are used to supplement the bifurcation diagrams from single-shot time-domain traces at these higher frequencies.

In the bent-pipe experiment (Fig 3(b)), the signal is generated by the source, travels through the circulator and then into one input port of a Wilkinson power combiner. The signal travels through the circulator through port 1 to port 2 and towards the diode. The signal then travels through a directional coupler, through some length of transmission line (length-1) and is reflected off the diode. The reflected wave passes back through the directional coupler and circulator in port 2 and out port 3 into another length of transmission line (length-2). The wave is then recombined with the driving signal and becomes the new incident wave via the power combiner. Some of the signal reflected by the diode is sampled (-10dB) and measured with a spectrum analyzer and oscilloscope. A concern for both experiments is the limited bandwidth of the microwave components and how this could suppress chaos. All components have had their s-parameters characterized over the full frequency range of the experiment.

The delay times of the experiments were estimated as follows. For the Partially Reflecting experiment, the length of the transmission line and $\varepsilon_r \sim 2$ (and $\varepsilon_r = \varepsilon/\varepsilon_0$ where $\varepsilon$ is the dielectric permittivity of the material inside the transmission line, and $\varepsilon_0$ is the permittivity of free space) were used to find the one-way delay time, and that time was recorded. For the Bent-Pipe experiment, the estimated round trip length of the resonator and $\varepsilon_r \sim 2$ were used to calculate the round trip delay times, and half that time was recorded as the relevant delay time scale $T$.

The packaged diodes are mounted on 50Ω BNC female-female adaptors. The leads are soldered directly onto the inner and outer conductors with the cathode end soldered to the outer conductor. Qualitatively similar results are obtained with the diode reversed. More background checks of the experiments were made by replacing the diode with a linear 3pF capacitor. In these checks, only period 1 behavior was found as a function of driving frequency and power.

## Experimental Results

Fig. 4(a) shows an experimental bifurcation diagram (taken from single-shot oscilloscope traces) of peak voltages as a function of driving power in dBm. This result was obtained using the bent-pipe design (Fig. 3b), a BAT41 diode, an 85 MHz driving signal, and a delay time of about 3.9 ns. Since the wavelength of the signal is about 2.4 meters ($\varepsilon_r$ ~2) and the length of the resonator is about 1.7 meters, the system is in the distributed regime. The driving power was varied from +16 dBm to +22.5 dBm.

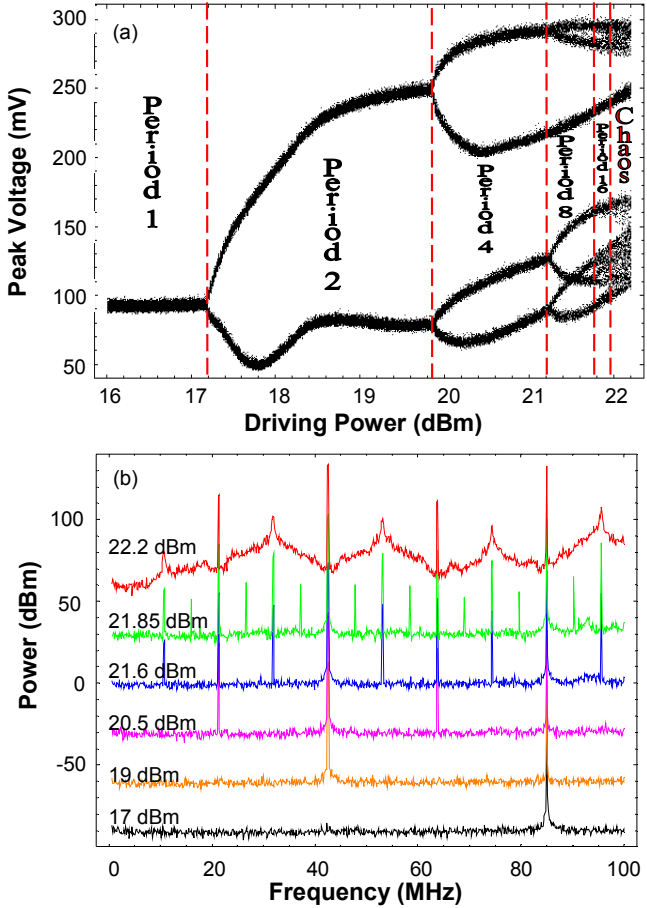

**Fig. 4: (a)** Experimental bifurcation diagram; BAT41 diode, $f$=85MHz, T~3.9ns, taken with the setup shown in Fig. 3b. The voltage signal is sampled by the oscilloscope through a -10dB directional coupler and a coaxial power splitter. **(b)** Corresponding experimental measurements of the frequency spectrum at selected driving powers. Each spectrum is offset by 30dB from the one below it.

As the driving power is increased, there are distinct points where period doublings occur. Clear regions of periods 1, 2, 4 and 8 can be seen through these splittings. The system noise and the accuracy limitations of the oscilloscope prevent a clear picture of period 8 and 16 behavior here; however, information on the type of orbit can be obtained by spectrum analysis. The frequency-power spectrum is shown in Fig. 4(b) to clearly illustrate periods 1, 2, 4, 8, 16 and chaos associated with the varying driving voltages in the period doubling cascade. Note, for example, we see period-2 behavior in the bifurcation diagram (Fig. 4(a)) at 19 dBm driving power. The spectrum in Fig. 4(b) of the system driven at 19 dBm shows the primary driving signal (85 MHz) as well as a signal at half the primary driving frequency. For each period doubling, another signal at $f_0/2^n$, where $f_0$ is the driving frequency and $n$ is the number of period doublings, appears along with its overtones. The period 16 peak is absent due to bandwidth limitations of the system components, mainly the directional coupler. We can still recognize the period 16 behavior from the overtones which occur at all integer multiples of $f_0/16$. We see that for chaos, the entire noise floor rises by at least 10 dB.

Although the bifurcation diagram obtained from numerical calculation of our model (see Fig. 2) appears to yield more complex behavior than that seen in the corresponding experimental bifurcation diagram (see Fig. 4), there are, nevertheless, some key qualitative similarities between the two. The ranges of powers for both diagrams are roughly the same, and both diagrams reach chaotic behavior at roughly +21 dBm. These similarities are representative of the parameters chosen to simulate the diode voltage-capacitance curve. $V_f$ was chosen to be 0.6 V, which is similar to the contact potential of a standard silicon p-n junction. $C_r$ was chosen to be 1 pF, which is slightly less than the measured junction capacitance of the BAT41 diode (4.6 pF) used in the experiment.

An experimental phase diagram resulting from the Partially Reflecting experiment (Fig. 3(a)) is shown in Fig. 5(b). In this case a 1N4148 diode was used with a transmission line of length of 1.83m and $\varepsilon_r$ ~ 2 (delay time about 8.6ns). The black dots represent observations of period 2, or more complicated, behavior as determined by spectrum analyzer data. Also shown is the experimental measurement of the small-signal system reactance as seen from the source looking into the circuit. A direct measurement of the small-signal (linear) reactance of the diode / transmission line system was made using an Agilent 4396B network / impedance analyzer, and then an electrical delay of -0.7m was added during analysis account for the other system components between the source and the transmission line (circulator, directional coupler etc.). Compared to the model predictions in Fig. 5(a), we see a strong similarity in both the phase diagrams and reactance vs. frequency. It seems that when the source sees a positive reactance looking into the circuit, period doubling and chaos are more likely to occur at lower driving amplitudes. The numerical phase diagram (Fig. 5(a)) was run with parameter values similar to the experimental setup (Fig. 5(b)) and the 1N4148 diode used in this experiment. The delay time used in the numerical analysis was the sum of the one way delay time of the transmission line (1.83m, $\varepsilon_r$ ~2) and the corresponding delay time from the estimated electrical delay (-0.7m, $\varepsilon_r$ =1).

Further experiments were performed with varying transmission line lengths (delay time $T$). As shown in Fig. 6(b), when the transmission line length was doubled to 3.66m (delay time about 17.3ns), it was seen that both the reactance and regions of period 2, or more complicated behavior, nearly doubled their frequency of occurrence. The electrical delay remained at -0.7m in the reactance analysis. Through numerical simulations, very similar changes in the phase diagrams can be seen. Fig. 6(a) is the corresponding numerical phase diagram for the experimental result of Fig. 6(b). When the delay time of the calculation that produced Fig. 5(a) was increased by $1.83\sqrt{2}/c \approx 8.6 ns$ (corresponding to the added transmission line length in the experiment), the number of black regions in the 400-800 MHz range nearly doubled and agreed accurately with the experiment. The theoretical reactance curves are also shown and behave in the same way, agreeing with the black regions of the phase diagrams and with the experimental reactance curves. Experiments of the BAT86 and NTE519 diodes run under the same conditions displayed very similar results; however, the phase diagrams of each diode showed different values for the onset of period doubling. The 1N4148 shows a period doubling transition of ~20dBm, as seen in Fig. 5(b) and 6(b), as opposed to ~25dBm in the NTE519 case and ~35dBm in the BAT86 case.

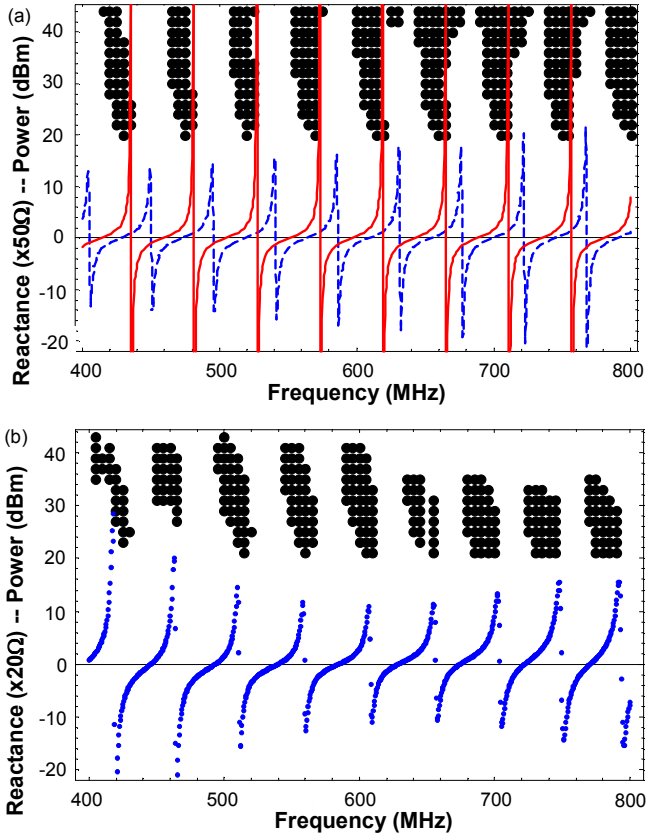

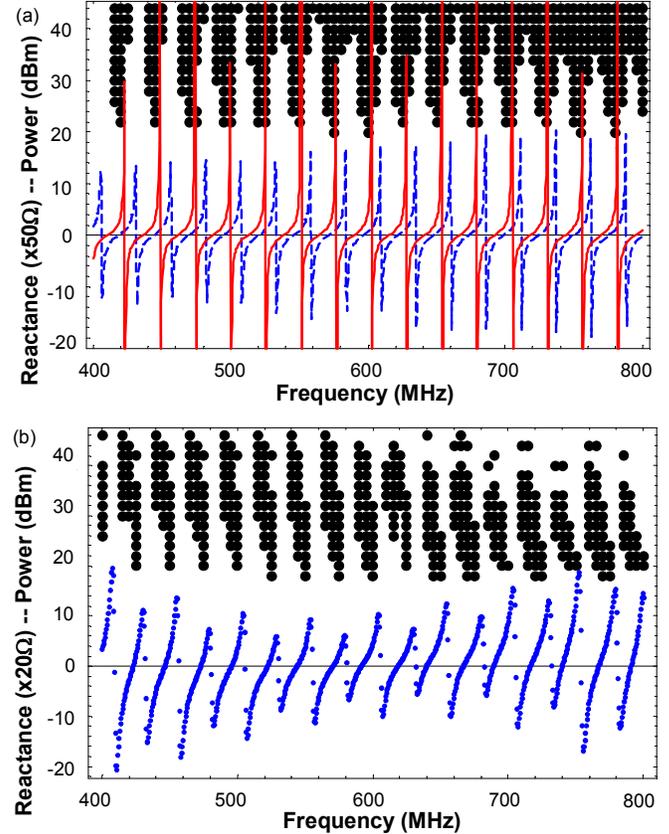

**Fig. 5:** (a) Numerical phase diagram as a function of frequency for Fig. 1. Black dots represent observations of period-2 or more complicated behavior. Also shown are theoretical reactance curves for the diode under forward (red) and reverse (blue, dashed) capacitance conditions. (b) Experimental phase diagram of a 1N4148 diode on a 1.83m transmission line using the Partially-Reflecting experiment (Fig. 3(a)). The lower curve is an experimental measurement of the system reactance as seen by the source.

**Fig. 6:** (a) Numerical phase diagram as in Fig. 5(a), but T has been increased by 8.6 ns. (b) Experimental phase diagram as in Fig. 5(b), but T has been increased by adding an additional 1.83m, 75Ω transmission line. This corresponds to the change in T of Fig. 6(a).

DC voltage biasing was also incorporated into the bent-pipe experiment in order to extend the driving power range over which chaos is observed. A DC voltage applied across the diode has two major effects on this experiment:[10,11] 1) it alters the reverse recovery time of the diode, 2) there is a shift of the nonlinear capacitance-voltage curve. By increasing the forward bias, the diode will be shifted into the strongly nonlinear $C(V)$ regime, and by applying a reverse bias, the diode shifts into the more linear regime. It was found that by applying a forward bias across the diode, period doubling and chaos could be seen for certain frequencies and powers where no period doubling or chaos was seen under zero bias. We were also able to shift period doubling and chaos to lower driving powers under varying DC bias. By applying a reverse bias, period doubling and chaos could be turned into period 1 behavior. In Fig. 7, with an 875 MHz driving signal, we see that as DC bias is increased, a period doubling transition appears where none existed before under 0 or negative bias.

## Discussion

There are several important relationships that play a role in generating non-trivial dynamical behavior in this system. First, from both the experimental and numerical phase diagrams, we see that period doubling and chaos occur when there is a positive value for the reactance as seen by the source. If we compare this system to the driven, lumped element RLD circuit, we see that the distributed system has a lumped nonlinear capacitor (diode), some resistance, yet no inductor. When the system obtains a positive value of reactance (at a given frequency and delay time), in effect it is like adding an inductive energy storage mechanism (or degree of freedom) to the problem and this system becomes qualitatively similar to the driven, lumped RLD circuit with the possibility of resonance and chaos.

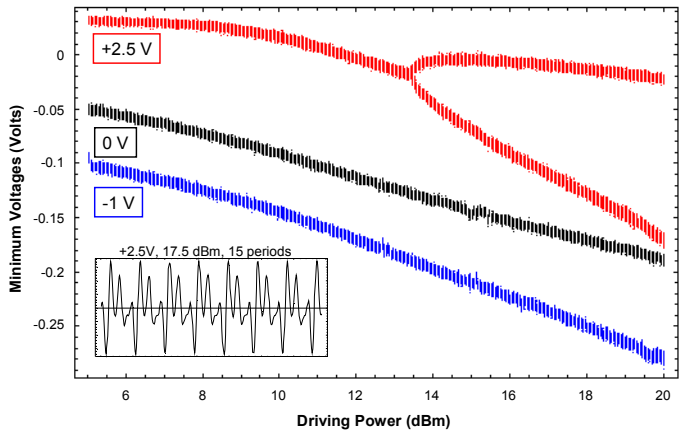

**Fig. 7:** Experimental bifurcation diagrams taken at different DC bias voltage values across the diode using the Bent-Pipe experiment (Fig. 3(b)). $f$=875 MHz, 1N4148 diode, and T~4.1ns. By applying a forward bias across the diode, a period doubling transition is revealed. A sample of the period-2 waveform of the +2.5V case driven at +17.5 dBm is inset. The -1 volt bifurcation diagram is offset by -0.05 volts and the +2.5 volt case is offset by +0.05 volts.

Table II summarizes the results of experiments done for this work. Some general observations can be made about when period doubling and chaos occur. We see that the 1N4148, BAT86, BAT41 and NTE519 diodes exhibit either period doubling, chaos, or both, whereas the NTE588, MV209, 5082-2835, and 5082-3081 diodes never show period doubling or chaos for the listed experimental conditions. First, one can note that only diodes with a reverse recovery time (~ 4 ns) comparable to the delay times $T$ of the experiments show period doubling and chaos. Second, we note that for these same diodes, the driving periods ($1/f$) used were comparable to the reverse recovery times. For instance, a recovery time of 4 ns corresponds to a frequency of 250 MHz, which is on the order of the driving frequencies used in these experiments. All diodes with $\tau_{RR} >> 1/f$ and $T$ did not show period doubling or chaos. Specifically, in the instances of the bent-pipe experiments, the delay times are between 3 and 7 ns. The diodes that show both period doubling and chaos in the bent-pipe experiment (1N4148, BAT41 and NTE519) all have reverse recovery times rated at 4 to 5 ns by the manufacturers. For reverse recovery times much longer than the delay times (as with the NTE588, MV209, and 5082-3081 diodes), no period doubling or chaos occur under the same set of experimental conditions.

The highest driving frequency at which chaos has been observed was 1.105GHz (Fig 8). The experiment used the bent-pipe design, a delay time of ~3.5ns, the NTE519 diode, 0 volts of DC biasing and about +50 dBm of driving power. Fig. 8 shows the experimental frequency spectrum of this case plotted along with the noise floor.

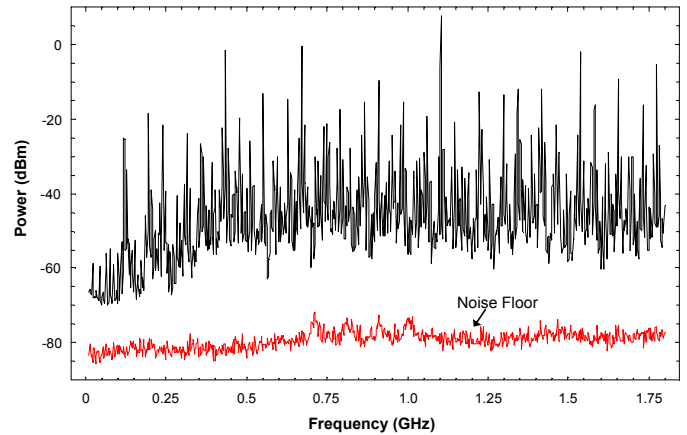

**Fig. 8:** Frequency spectrum of chaos at the highest driving frequency at which it was found, compared to the noise floor. System parameters: $f$=1.105GHz, NTE519 diode, bent-pipe experiment with bias tee, T~3.5ns, ~+50dBm driving power, 0V DC bias.

## Conclusions

We have introduced a new model of chaos in a driven distributed transmission line / diode system. This model qualitatively represents a means to generate chaos at microwave frequencies, for example in computer interconnects with ESD diodes protecting integrated circuits. We have examined numerical predictions of chaos. The theory predicts period doubling and chaos, and the model results have given us some indication of what conditions will most likely induce period doubling and chaos in this system. The model and experimental phase diagrams show that period doubling, or more complicated behavior, occurs when the source sees a positive small-signal reactance looking into the circuit. From the table of experimental results (Table II), it appears that period doubling and chaos occur when the important time scales (reverse recovery time of the diode, delay time of the system, and driving period) are roughly on the same order ($\tau_{RR} \sim T \sim 1/f$). DC biasing (which may also result from rectification[11]) was also shown to be a useful parameter controlling the occurrence of chaos in the system. The Partially-Reflecting experiment has shown period doubling and the Bent-Pipe experiment has shown period doubling and chaos. Both experiments have shown these results in high driving frequency, distributed situations. The highest driving frequency at which we have seen chaos is 1.105GHz.

## Acknowledgements

We thank S. Hemmady and D. Mircea for their support on this study and R. Magda for pointing out Ref. [23]. We also thank Agilent, LeCroy and Tektronix for loaning oscilloscopes used in this work. This work was supported by the DOD MURI for the study of microwave effects on electronics under AFOSR Grant F496200110374 and AFOSR DURIP Grants FA95500410295 and FA95500510240.

**Table II: Summary of Experimental Results**

Measurements of the reverse recovery times were made as described in reference 10. Square wave pulses were applied to a resistor-diode circuit using an HP 33120A function generator. The pulses were 2μs long, 1s apart, 3 volts peak-to-peak, and 0 volts dc offset. R = 24 Ω. The values in the table with an * represent values that could not be measured directly, therefore the manufacturer's value is listed instead. Measurements of $C_j(0)$ were made using an Agilent 4285A LCR meter in Cp-Rp mode at 1 MHz, 50mV AC and 0V DC. PD represents "period doubling."

| Diode | $\tau_{RR}$ (ns) | $C_j(0)$ (pF) | Experiment | Delay Time T (ns)[i] | Result | Min. Pow. to PD | ~ f range for Result |
|---|---|---|---|---|---|---|---|
| 1N4148 | 4* | 0.7 | Partially Reflecting | 8.6, 17.3 | **PD** (Fig. 5b, 6b) | ~18 dBm | 0.4–1.0 GHz periodically |
| | | | Bent-Pipe | 3.0, 3.5, 3.9, 4.1, 4.4, 5.5, 7.0 | **PD**, **Chaos**[ii] | ~14 dBm | 0.2–1.0 GHz (Fig. 7) |
| BAT86 | 4* | 11.5 | Partially Reflecting | 8.6, 17.3 | **PD** | ~ 35 dBm | 0.4–1.0 GHz periodically |
| | | | Bent-Pipe | 3.0, 3.5, 3.9, 4.1, 4.4, 5.5, 7.0 | Per 1 only | --- | 20-800 MHz |
| BAT41 | 5* | 4.6 | Partially Reflecting | 8.6, 17.3 | Per 1 only | --- | 0.4-1.0 GHz |
| | | | Bent-Pipe | 3.9 | **PD**, **Chaos** | ~ 25 dBm | 43 MHz |
| | | | | | | ~ 17 dBm | 85 MHz (Fig. 4) |
| | | | | 3.0, 3.5, 4.1, 4.4, 5.5, 7.0 | Per 1 only | --- | 20-800 MHz |
| NTE519 | 4* | 1.1 | Partially Reflecting | 8.6, 17.3 | **PD** | ~ 25 dBm | 0.4-1.0 GHz periodically |
| | | | Bent-Pipe | 3.0, 3.5, 3.9, 4.1, 4.4, 5.5, 7.0 | **PD**, **Chaos**[ii] | ~16 dBm | 0.5-1.2 GHz (Fig. 8) |
| NTE588 | 35 | 116 | Partially Reflecting | 8.6, 17.3 | Per 1 only | --- | 0.02 - 1.2 GHz |
| | | | Bent-Pipe | 3.0, 3.5, 3.9, 4.1, 4.4, 5.5, 7.0 | | | |
| MV209 | 30 | 66.6 | Partially Reflecting | 8.6, 17.3 | Per 1 only | --- | 0.02 - 1.2 GHz |
| | | | Bent-Pipe | 3.0, 3.5, 3.9, 4.1, 4.4, 5.5, 7.0 | | | |
| 5082-2835 | <15 | 0.7 | Partially Reflecting | 8.6, 17.3 | Per 1 only | --- | 0.02 - 1.2 GHz |
| | | | Bent-Pipe | 3.0, 3.5, 3.9, 4.1, 4.4, 5.5, 7.0 | | | |
| 5082-3081 | 380 | 2.0 | Partially Reflecting | 8.6, 17.3 | Per 1 only | --- | 0.02 - 1.2 GHz |
| | | | Bent-Pipe | 3.0, 3.5, 3.9, 4.1, 4.4, 5.5, 7.0 | | | |

[i] Delay times are one-way for the simple experiment and ½ round trip for the Bent-Pipe experiment.
[ii] With dc bias applied to the diode.